\documentclass{aa} 
\usepackage{graphicx}
\usepackage{lscape}
\usepackage{longtable}
\usepackage{arydshln}
\usepackage{amsmath}
\usepackage{txfonts}
\usepackage{natbib}
\usepackage[htt]{hyphenat}
\usepackage{breqn}
\usepackage[normalem]{ulem}

\usepackage[dvipsnames]{xcolor}

\bibpunct{(}{)}{;}{a}{}{,} 

\begin{document}

\title{Observations of sulfuretted species in HL Tau} 

\author{P. Rivi\`ere-Marichalar \inst{1}
\and R. le Gal \inst{2,3}
\and A. Fuente    \inst{4}
\and D. Semenov \inst{5,6}
\and G. Esplugues \inst{1}
\and D. Navarro-Almaida    \inst{4}
\and S. Facchini \inst{7}
}

\institute{Observatorio Astron\'omico Nacional (OAN,IGN), Calle Alfonso XII, 3. 28014 Madrid, Spain 
                   \email{p.riviere@oan.es}
        \and Institut de Radioastronomie Millim\'etrique, 300 rue de la Piscine, F-38406 Saint-Martin d'H\`eres, France 
        \and Institut de Plan\'etologie et d'Astrophysique de Grenoble (IPAG), Universit\'e Grenoble Alpes, CNRS, 38000 Grenoble, France 
        \and  Centro de Astrobiolog\'ia (CAB), INTA-CSIC, Carretera de Ajalvir Km. 4, Torrej\'on de Ardoz, 28850 Madrid, Spain
        \and Zentrum f\"{u}r Astronomie der Universit\"{a}t Heidelberg, Institut f\"{u}r Theoretische Astrophysik, Albert-Ueberle-Str. 2, 69120 Heidelberg, Germany 
        \and Max-Planck-Institut f\"{u}r Astronomie, K\"{o}nigstuhl 17, 69117 Heidelberg, Germany 
        \and Dipartimento di Fisica, Universit\`a degli Studi di Milano, Via Celoria 16, 20133 Milano, Italy}

\authorrunning{Rivi\`ere-Marichalar et al.}
\titlerunning{Observations of sulfuretted species in HL Tau}
\date{}

 \abstract 
{Protoplanetary disks inherit their chemical composition from their natal molecular cloud, although the extent to which this material is preserved versus reset through chemical reprocessing remains an open question. Understanding this balance is a major topic in astrochemistry and star and planet formation. Comparing the chemical composition of the envelope and the protoplanetary disk is key to solving the topic. However, disentangling protoplanetary disk emission from envelope emission is not an easy task.}
{The goal of this paper is to investigate the chemical differences between the disk and the surrounding envelope by comparing the column density ratios of a few selected species in each region. The source we focus on is HL Tau, where molecular absorption lines from the envelope have been detected, thus allowing for the derivation of column densities and molecular abundances.}
{We present new NOEMA observations of HL Tau targeting the following species: CS, H$\rm _2$CO,  H$\rm _2$S, and SO$\rm _2$. We produced zeroth-, first-, and second-moment maps for the species where emission was detected and used them to analyze the spatial distribution and kinematic properties of the different molecules in the disk and the envelope. We derived the column densities and compared the values derived for the envelope and disk. We also computed the rotational diagram for the SO$\rm _2$ detected transitions.}
{Assuming two different temperature regimes, 17 and 58 K, we derived column densities for the species surveyed in the disk and compared them with values derived for the envelope. We find large differences in the derived column density ratios of the surveyed molecules, especially for N(CS)/N(H$\rm _2$S), which is 40 to 50 times larger in the envelope. We attribute these variations to the different excitation and UV-irradiation regimes in the disk and envelope. We also note strong gradients in the ratios between different positions of the disk and tentatively attribute them to different levels of turbulence at different azimuths. }
{The observed differences in molecular ratios in the envelope and the disk are suggestive of chemical reprocessing of the gas during the formation and evolution of the protoplanetary disk.} 


\maketitle

\section{Introduction} \label{Introduction}
Protoplanetary disks are the birthplace of planets \citep{Williams2011}. As such, they set the initial chemical abundances available for atmosphere formation \citep{Cridland2016}. Protoplanetary disks inherit their chemical composition from the molecular cloud where they are born \citep[see, e.g., ][]{Oberg2023}. The chemical composition inherited from the molecular cloud, however, is subject to a level of reprocessing that remains uncertain, but comparing the chemical composition of the protoplanetary disk and the stellar envelope provides a means to study this topic since the composition of gas in the envelope should be close to that in the molecular cloud. This is not an easy task, though, since separating the contribution from the envelope from that of the disk is a complex issue and usually relies on detailed modeling of the system chemistry, where assumptions and uncertainties in the model parameters can jeopardize the analysis of the results. 
\begin{table*}[]
\caption{Line properties and fluxes.}
\label{Tab:line_fluxes}
\centering
\begin{tabular}{lllllll}
\hline \hline
Species & Transition & $\nu_0$ & Beam size& $\rm E_u$ & A$\rm _{ij}$ & Flux \\
(--) & (--) & (GHz) & (\arcsec) & (K) & (s$\rm ^{-1}$) & Jy km s$\rm ^{-1}$ \\
\hline
SO$\rm _2$ & 4$\rm _{2,2}$-4$\rm _{1,3}$ & 146.605 & 0.81 $\times$ 0.31 & 19.02892 & 2.4697$\rm \times 10^{-5}$ & 0.028$\rm \pm$0.005 \\
CS & 3-2 &  146.969 & 0.81 $\times$ 0.31 & 14.10692 & 6.0517$\rm \times 10^{-5}$  & 0.35$\rm \pm$0.05 \\
H$\rm _2$CO & 2$\rm _{1,1}$-1$\rm _{1,0}$ & 150.498 & 0.81 $\times$ 0.30 & 22.61771 & 3.5950$\rm \times 10^{-5}$ & 0.26$\rm \pm$0.03\\
SO$\rm _2$ & 2$\rm _{2,0}$-2$\rm _{1,1}$ & 151.379 & 0.74 $\times$ 0.27 & 12.58483 & 1.8752$\rm \times 10^{-5}$ & 0.014$\rm \pm$0.003\\ 
SO$\rm _2$ & 5$\rm _{2,4}$-5$\rm _{1,5}$ & 165.144 & 0.74 $\times$ 0.27 & 23.58866 & 3.1223$\rm \times 10^{-5}$ & 0.035$\rm \pm$0.06 \\
H$\rm _2$S & 1$\rm _{1,0}$-1$\rm _{0,1}$ & 168.762 & 0.72 $\times$ 0.26 & 27.87738 & 2.6773$\rm \times 10^{-5}$ & 0.27$\rm \pm$0.03 \\
SO$\rm _2$ & 14$\rm _{1,13}$-14$\rm _{0,14}$ & 163.605 & 0.75 $\times$ 0.27 & 101.75216 & 3.0054$\rm \times 10^{-5}$ & 0.026$\rm \pm$0.005\\
\hline
SO$\rm _2$ & 15$\rm _{5,11}$-16$\rm _{4,12}$ & 150.381 & 0.81 $\times$ 0.30 & 171.68074 & 6.9523$\rm \times 10^{-6}$ & $\rm <$0.026 \\
SO$\rm _2$ & 18$\rm _{2,16}$-17$\rm _{3,15}$ & 163.119 & 0.75 $\times$ 0.27 & 170.76008 & 1.3540$\rm \times 10^{-5}$ & $\rm <$0.035\\
SO$\rm _2$ & 24$\rm _{2,25}$-25$\rm _{1,25}$ & 168.790 & 0.72 $\times$ 0.26  & 292.73681 & 5.4519$\rm \times 10^{-5}$ & $\rm <$0.039\\
\hline
\end{tabular}
\tablefoot{Molecular data from CDMS \citep{Muller2005}. Line fluxes were computed using Keplerian-masked integrated intensity maps (see Sect. \ref{Subsec:CS,H2S,H2CO}). Flux uncertainties are the quadratic sum of the propagated rms and the calibration uncertainty (10\%).}
\end{table*}

HL Tau is a well-known Class I-II \citep{Furlan2008} protostar in Taurus. The system harbors a protoplanetary disk where a series of concentric rings have been observed in the submillimeter band with the Atacama Large Millimeter Array (ALMA) \citep{ALMA2015} and is also embedded in an infalling and rotating envelope \citep{Hayashi1993}. The concentric rings observed in the continuum have been associated with the presence of forming gas giant planets \citep[see e. g.][]{Dipierro2015, Dong2015, Kanagawa2015}, but there are alternative explanations, such as dust growth \citep{Zhang2015} and gravitational instabilities \citep{Booth2020}. High angular resolution observations by \cite{Yen2016} with ALMA have revealed hints of gas cavities at radii 28 and 69 au, with the inner one coinciding with the dust gap at 32 au, and the outer one located in the bright continuum ring at 69 au overlaps with the millimeter continuum gaps at 64 and 74 au. Further observations with ALMA of $\rm ^{13}$CO and C$\rm ^{18}$O have revealed the presence of arc-like structures connected to the central regions of HL Tau extending out to as much as 2000 au with strong velocity gradients along them \citep{Yen2017}, likely tracing infalling material. Observations of HCO$\rm ^+$ also have revealed the presence of a spiral structure that is most likely infalling onto the disk \citep{Yen2019}. Thus, HL Tau seems to be actively interacting with its surroundings \citep{Wu2018}. Large-scale interactions of protostellar systems with their surroundings have become a hot topic in the past few years \citep{Pineda2020, Pineda2023}.  Understanding the impact of large-scale streamers in the evolution of the disk is paramount to understanding mass buildup. The streamers will also have a dynamic influence on the disk and will likely modify the chemical composition by bringing fresh material to the disk. In \cite{Yen2019}, the authors further studied the Keplerian profile of the molecular emission, concluding that HCO$\rm ^+$ 3-2 is tracing material in the disk, and they computed a stellar mass of 2.1 M$\rm _\odot$. They also detected absorption in their observations, possibly caused by an infalling envelope. Thus, HL Tau is the perfect laboratory to compare the chemical composition of the protoplanetary disk and the envelope. 

In this paper, we present millimeter observations of HL Tau performed using NOEMA. Our observations detect the deep absorption observed by \cite{Yen2019} in the three species detected -- CS, H$\rm _2$S, and H$\rm _2$CO -- thus allowing us to study the chemical budget in the envelope. We present our observations and describe the data reduction process in Sect. \ref{Sect:obs_data_red}. In Sect. \ref{Sect:results} we present our results. In Sect. \ref{Sect:discussion} we discuss the implications of our results, and in Sect. \ref{Sect:summary} we provide a summary of the paper.

\begin{figure}[t!]
\begin{center}
  \includegraphics[width=0.45\textwidth]{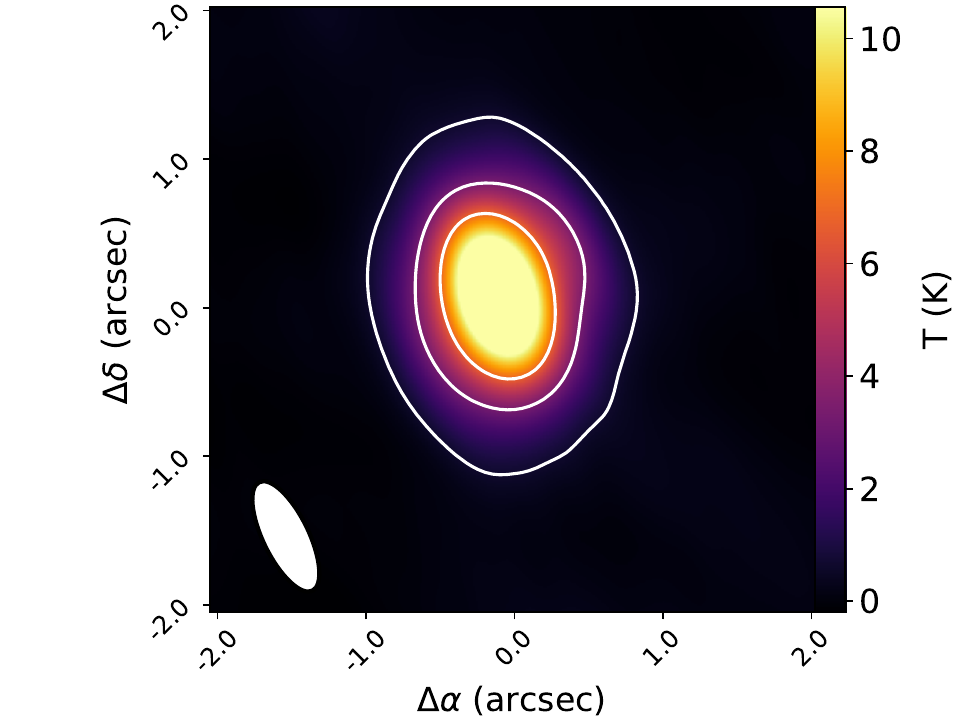}\\
  \caption{Continuum map at 2 mm. Contour levels are 5, 25, and 50 times the rms.}
 \label{Fig:cont_map}
\end{center}
\end{figure}

\begin{figure*}[t!]
\begin{center}
  \includegraphics[width=0.9\textwidth, trim=0mm 10mm 0mm 0mm, clip]{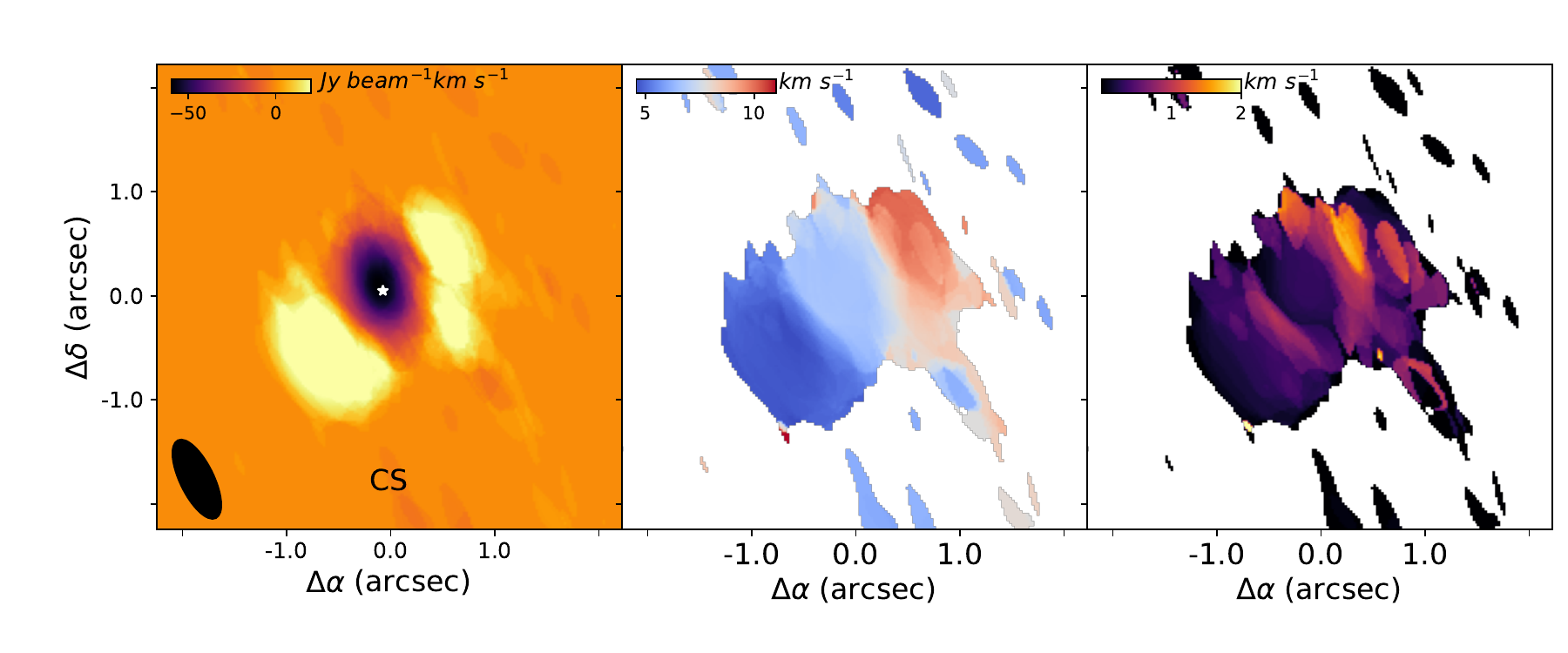}\\
  \includegraphics[width=0.9\textwidth, trim=0mm 10mm 0mm 0mm, clip]{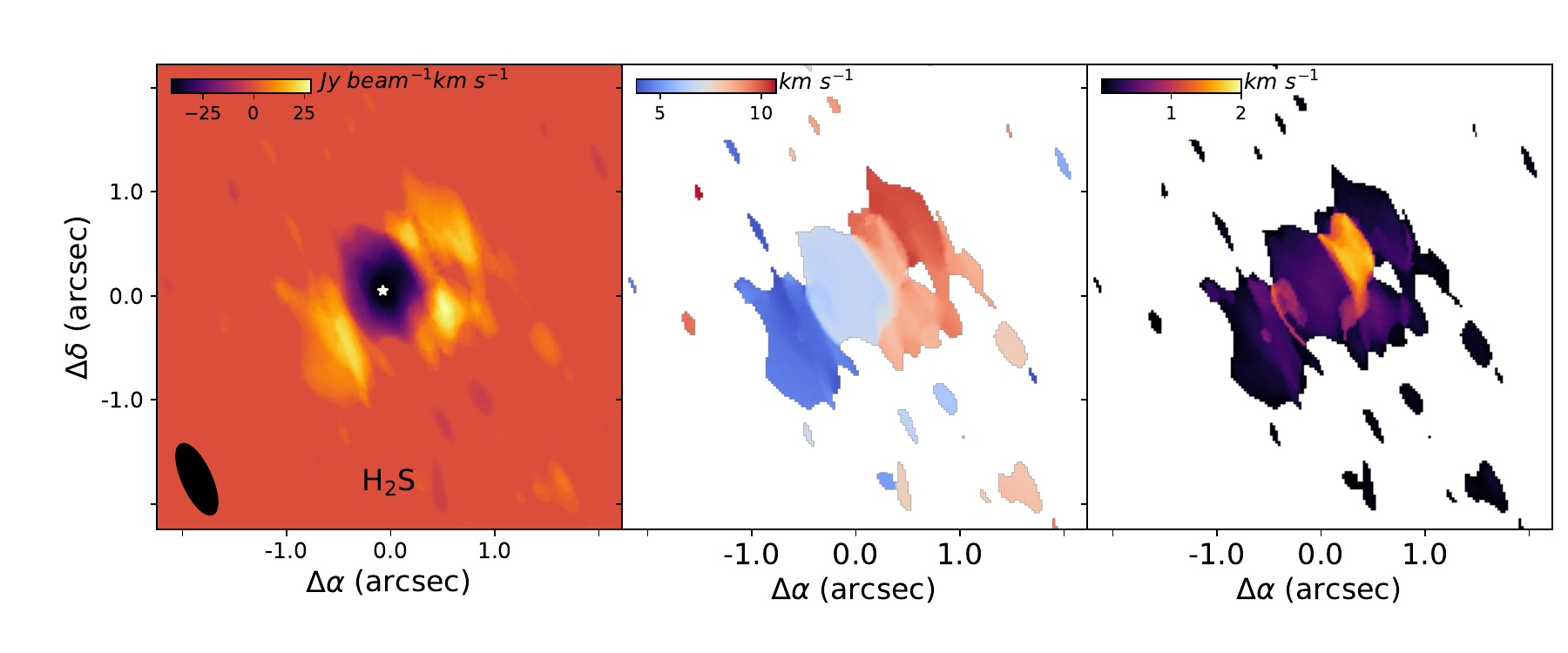}\\
  \includegraphics[width=0.9\textwidth, trim=0mm 10mm 0mm 0mm, clip]{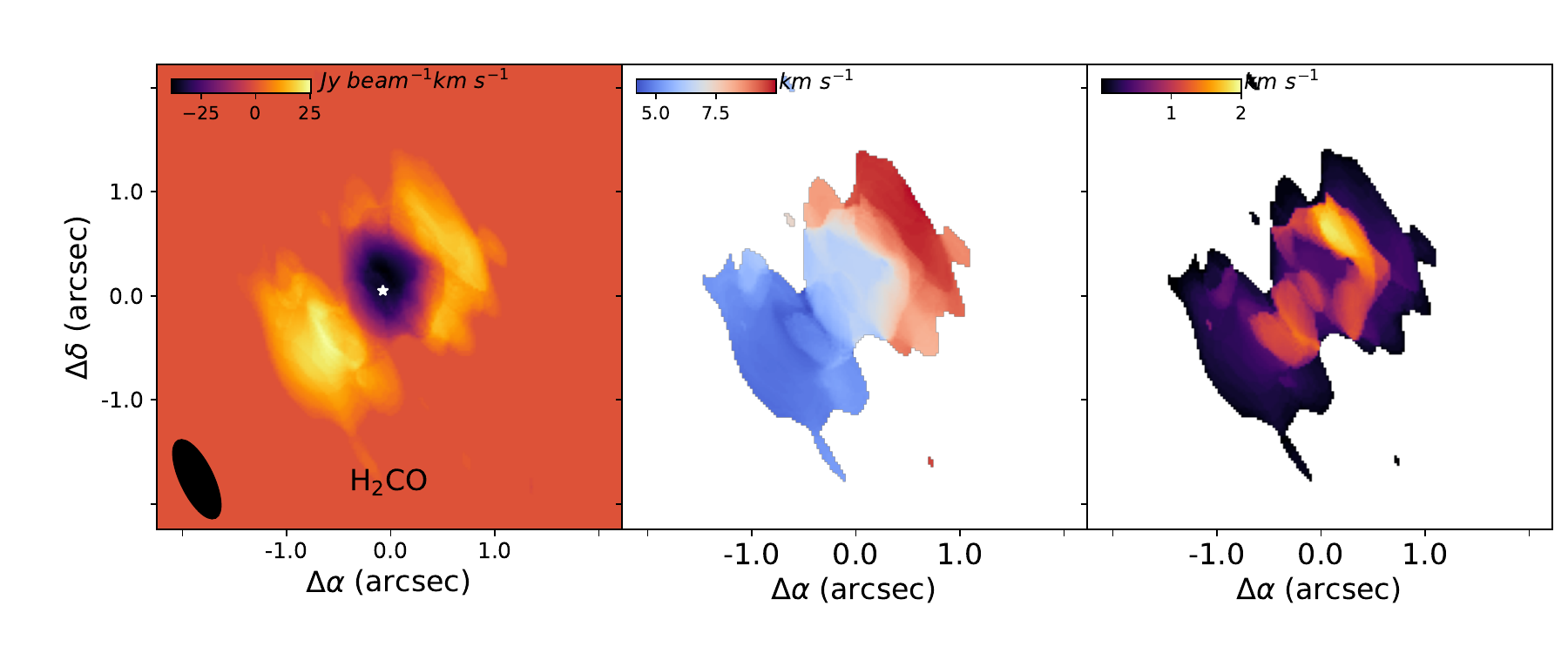}\\
  \caption{From left to right: Continuum-subtracted zeroth-, first-, and second-moment map of CS 3-2 (top), H$\rm _2$S   1$\rm _{10}$-1$\rm _{01}$ (middle), and H$\rm _2$CO $\rm 5_6-4_5$ (bottom) observed with NOEMA. Before the moment calculation, a four-sigma clipping mask was applied to the data cubes. The cubes were smoothed by a factor of three before computing the moment maps. Only channels in the range 3 to 11 km s$\rm ^{-1}$ and with a S/N larger than four have been used.}
 \label{Fig:moments}
\end{center}
\end{figure*}

\begin{figure*}[t!]
\begin{center}
  \includegraphics[width=0.9\textwidth, trim=0mm 10mm 0mm 0mm, clip]{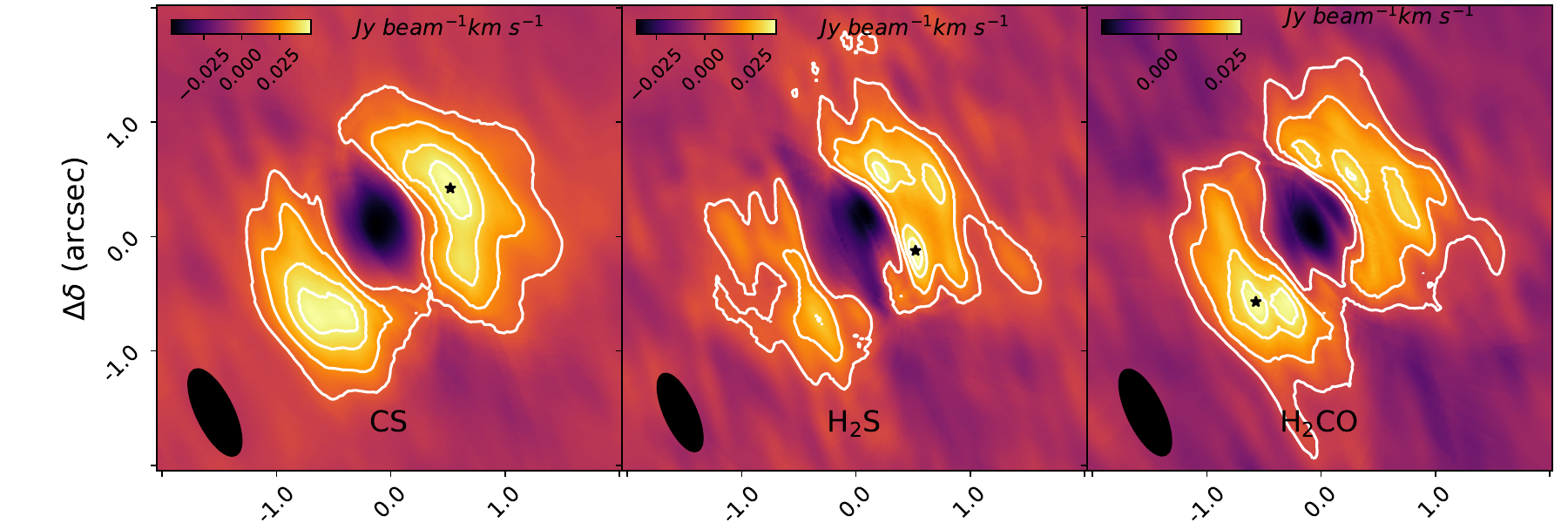}\\
  \caption{From left to right: Integrated intensity maps of CS 3-2 (left), H$\rm _2$S 1$\rm _{10}$-1$\rm _{01}$ (middle), and H$\rm _2$CO $\rm 5_6-4_5$ (right) after a Keplerian mask has been applied. The black star in each map marks the position of the emission peak of the map.}
 \label{Fig:moments_KeplerMask}
\end{center}
\end{figure*}

\section{Observations and data reduction}\label{Sect:obs_data_red}

Observations were carried out with NOEMA as part of project W22BA (PI: P. Rivi\`ere-Marichalar), combining the A- and B-array configurations in Band 2 from February 15 and March 22, 2023, with projected baselines ranging from 32.0 m ( 17.6 k$\lambda$) to 1664.0 m ( 917.5 k$\lambda$). The phase tracking center of our observation was $\alpha$(J2000) = 04h31m38s4.511, $\beta$(J2000) = 18$^{\circ}$13\arcmin57.\arcsec860. The A-configuration observations were observed on February 15, 2023, under good weather conditions and with all twelve antennas, accumulating an on-source time of 3.7h. The atmospheric phase stability ranged between 15$^\circ$ and 50$^\circ$ RMS, and the precipitable water vapor values were between 1 and 3\,mm. The B-configuration observations were observed on March 22, 2023, under relatively poor weather conditions with 11 antennas, accumulating an on-source time of 3h. The atmospheric phase stability ranged between 30$^\circ$ and 100$^\circ$ RMS, and precipitable water vapor values were between 2 and 5\,mm. The 2SB NOEMA receivers were tuned to cover a total instantaneous nominal bandwidth of $\sim$15.5 GHz per polarization, from 146.4 to 154.2 GHz in the lower sideband and 161.8 to 169.6 GHz in the upper sideband (USB), with a spectral resolution of 2 MHz. The correlator PolyFiX was used in high-resolution mode, allowing the addition of 64 MHz spectral windows and providing 62.5 kHz channel resolution to specifically target the H$\rm _2$S 1$\rm _{1,0}$-1$\rm _{0,1}$ transition at 168.7 GHz and additional sulfuretted species such as the CS (3-2) transition at 146.969 GHz, seven SO$\rm _2$ transitions at different wavelengths (see Table \ref{Tab:line_fluxes}), and H2CO 2$\rm _{1,1}$-1$\rm _{1,0}$ at 150.498GHz.

The NOEMA data were calibrated using the continuum and line interferometer calibration package of GILDAS.\footnote{https://www.iram.fr/IRAMFR/GILDAS/} We used 3C 84 and 2013+370 as the bandpass calibrators. Amplitude and phase calibration were performed using J0440+146 and 0507+179 for both tracks, with flux densities bootstrapped to standard references LkH$\alpha$101, MWC 349, and 2010+723. The absolute flux calibration accuracy was estimated to be within 15\%. Imaging was performed with the MAPPING GILDAS package using natural weighting. The resulting beam sizes are provided in Table \ref{Tab:line_fluxes}.

\begin{figure}[t!]
\begin{center}
  \includegraphics[trim=7mm 0mm 0mm 0mm, clip, width=0.45\textwidth]{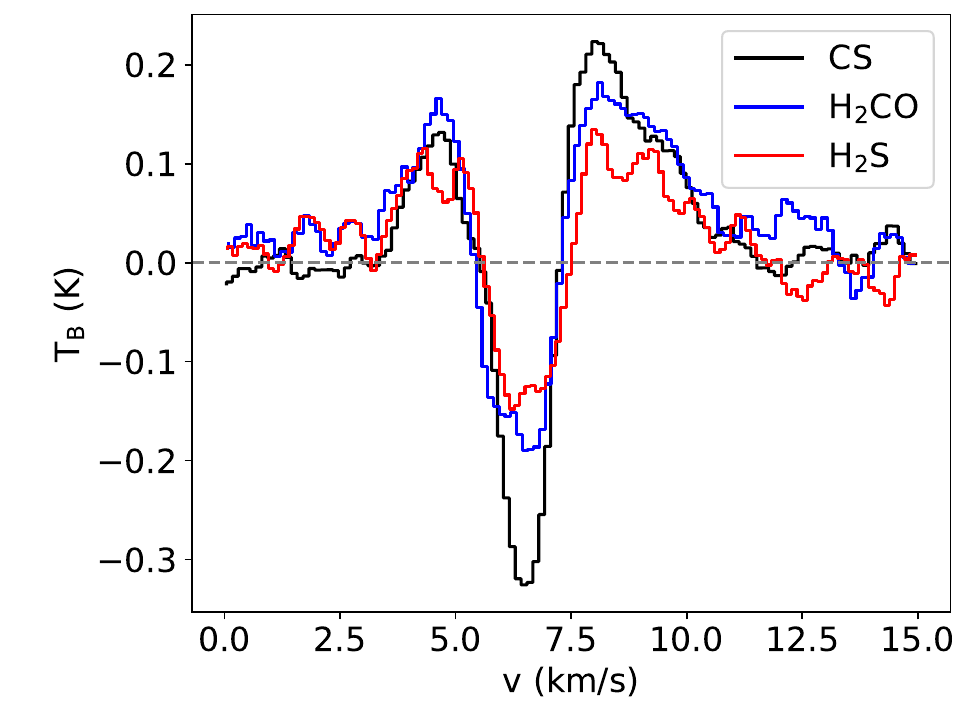}\\
  \caption{Continuum-subtracted average spectra of CS, H$\rm _2$CO, and H$\rm _2$S. The spectra shown have been integrated within a radius of 1.6$\arcsec$ from the center.}
 \label{Fig:integrated_spectra}
\end{center}
\end{figure}

\section{Results}\label{Sect:results}
In Fig. \ref{Fig:cont_map} we show the continuum map at 2 mm. Emission is observed until roughly 1.2$\arcsec$ from the center. We detected molecular emission from CS, H$\rm _2$CO, H$\rm _2$S, and SO$\rm _2$. In Table \ref{Tab:line_fluxes}, we provide the spectral line properties from the Cologne Database for Molecular Spectroscopy \citep[CDMS; ][]{Muller2005} and line fluxes for all the species detected and upper limits for non-detected species. We describe how line fluxes were computed in Sects. \ref{Subsec:CS,H2S,H2CO} and \ref{Subsec:SO2}. The observations can be divided into two groups. The first group shows a broken ring morphology surrounding a region of absorbed emission toward the center and includes CS, H$\rm _2$CO, and H$\rm _2$S. The second group, consisting of all the detected SO$\rm _2$ transitions, shows emission only on the western side of the disk, and in a separated lobe to the southwest. In the following, we analyze the results for each group separately. 

\subsection{CS, H$\rm _2$S, and H$\rm _2$CO}\label{Subsec:CS,H2S,H2CO}
We produced zeroth-, first-, and second-moment maps for all the species detected. The data cubes were spectrally smoothed by a factor of three before computing the moment maps to boost the S/N. In Fig. \ref{Fig:moments} we show the zeroth-, first-, and second-moment maps of CS, H$\rm _2$CO, and H$\rm _2$S. Only channels with an S/N larger than four were included. The three species show prominent absorption features toward the central star, as previously reported for other species such as HCO$\rm ^+$, CN, and HCN \citep{ALMA2015, Yen2019}. Rings of emission surround the central absorption. CS and H$\rm _2$CO show brighter emission toward the southeast, but for H$\rm _2$S, the brightest spot is toward the southwest. The CS 3-2 map shows an arc of extended emission toward the southwest, the counterpart of the HCO$^+$ 3-2 spiral arm observed in \cite{Yen2019}. The first-moment map of CS 3-2 shows hints of a velocity gradient along the structure, but observations at higher angular resolution are needed to confirm this. The first-moment maps show the characteristics of Keplerian rotation. The second-moment maps (intensity-weighted average velocity dispersion) show a widening of the emission lines toward the northwestern part of the disk. This kind of line widening is present in other systems where spiral arms perturb the disk \citep{Tang2012}. In the case of HL Tau, the line broadening seems to be connected to the spiral observed in CS and HCO$\rm ^+$ \citep{Yen2019}. The spiral arm is also marginally detected in the H$\rm _2$S 1$\rm _{1,0}$-1$\rm _{0,1}$ map. In Fig. \ref{Fig:moments_KeplerMask} we show the integrated intensity maps after applying a Keplerian mask computed using \texttt{Bettermoments} \citep{Teague2018}.  We used the \texttt{make mask} tool\footnote{https://github.com/richteague/keplerian\_mask} \citep{makeMask} to produce the Keplerian masks. Given the limited S/N of our observations, we decided to fix the position of the mask center to that of the phase center of the map. We assumed a stellar mass of 2.1 M$\rm _{\odot}$, and $i$=46.7$^{\circ}$, $PA$=138$^{\circ}$ \citep{Yen2019}.  We were also interested in studying the spatial distribution of non-Keplerian motions, but we did not detect them at a 3$\sigma$ level. Table \ref{Tab:line_fluxes} includes line fluxes for CS, H$\rm _2$S, and H$\rm _2$CO integrated inside a radius of 1.6$\arcsec$ around the center of the zeroth-moment Keplerian-masked maps. The regions affected by absorption were masked out, and no $\sigma$ clipping was applied to the maps before computing the line fluxes.

In Fig. \ref{Fig:integrated_spectra}, we show the integrated spectra of the three species detected. The spectra were integrated within a radius of 1.6$\arcsec$ from the center. A strong absorption feature can be observed toward the center in all three spectra, but with varying intensities relative to the emission detected in the wings. The largest dynamical range is observed in the CS spectra (0.55 K), and the smallest is that of H$\rm _2$S (0.28 K), with H$\rm _2$CO lying in between (0.37 K).  

\subsection{Radial profiles}\label{Subsec:rad_prof}
\begin{figure}[t!]
\begin{center}
  \includegraphics[width=0.45\textwidth, trim=0mm 10mm 0mm 0mm, clip]{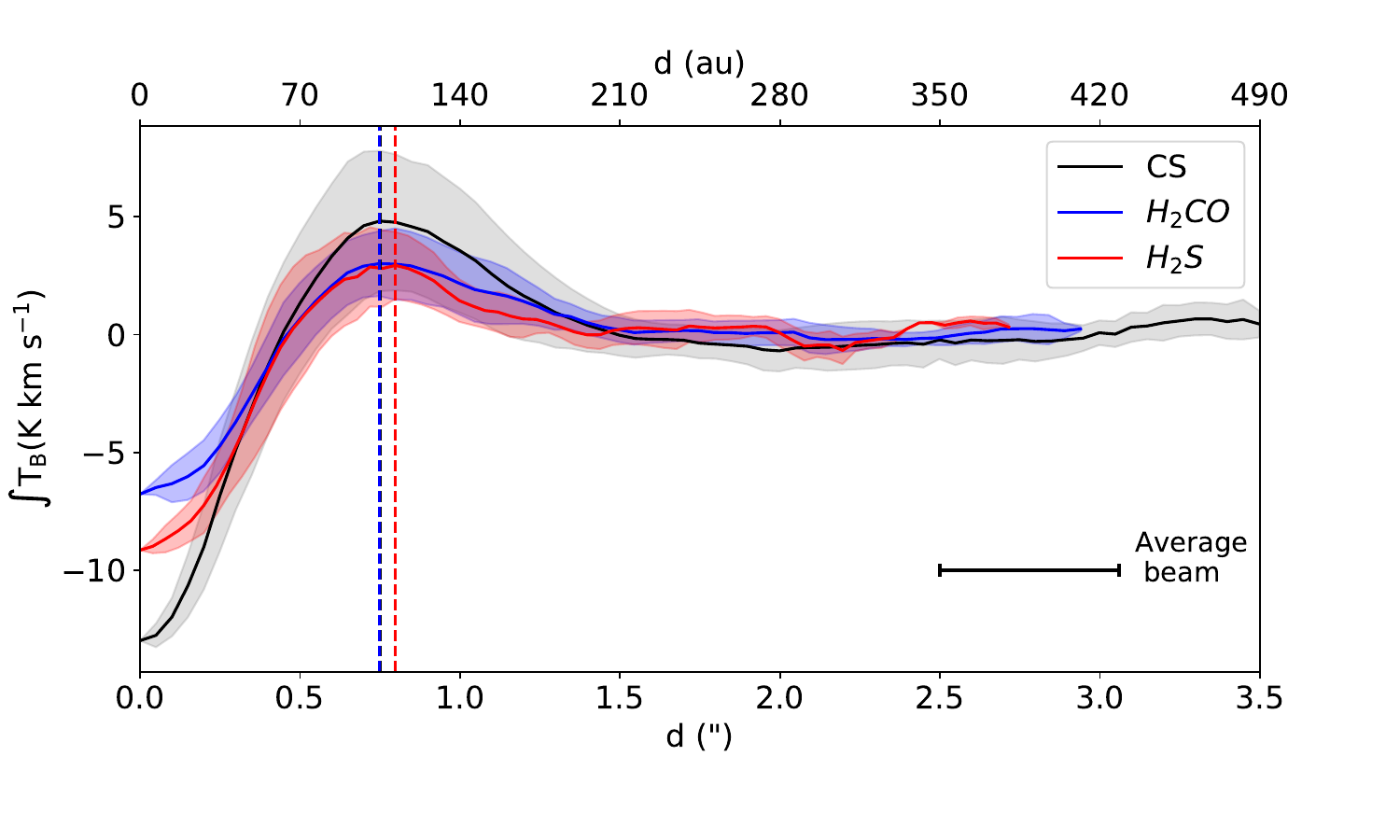}\\
  \caption{Radial profiles of the species detected in our survey. The radial profiles were computed on deprojected maps assuming i=46.7$^\circ$ and PA=138$^\circ$. The shaded area around each profile accounts for the standard deviation at each radial bin. The dashed vertical lines mark the position of the radial emission peaks. The black line in the bottom right corner shows the average beam size.}
 \label{Fig:rad_prof}
\end{center}
\end{figure}

In Fig. \ref{Fig:rad_prof} we show the azimuthally deprojected averaged radial profiles of the integrated intensity maps of CS, H$\rm _2$S, and H$\rm _2$CO with no Keplerian masks applied. We deprojected the maps before computing the radial profiles assuming $i$=46.7$^{\circ}$, $PA$=138$^{\circ}$ \citep{Yen2019}. The three species show very similar profiles. CS and H$\rm _2$CO peak at the same position (0.75$\arcsec$, $\sim$ 105 au), while H$\rm _2$S peaks slightly farther out  (0.8$\arcsec$, $\sim$ 111 au). The differences in the position of the peaks are smaller than the size of the beam, and therefore the three peaks are at the same distance within our angular resolution. Accordingly, we observed no hints of chemical segregation in the radial direction within the angular resolution of our observations. Line absorption is observed until 0.45\arcsec (63 au), roughly. Line emission is mostly detected at a maximum distance of 1.5\arcsec (210 au), although marginal emission seems present until 2\arcsec (280 au). H$\rm _2$S and H$\rm _2$CO show very similar profiles, reaching similar maximum azimuthally averaged intensities (3 K km s$\rm ^{-1}$, compared to 4.8 K km s$\rm ^{-1}$ for CS).

\subsection{SO$\rm _2$}\label{Subsec:SO2}
\begin{figure*}[t!]
\begin{center}
  \includegraphics[width=1.0\textwidth, clip]{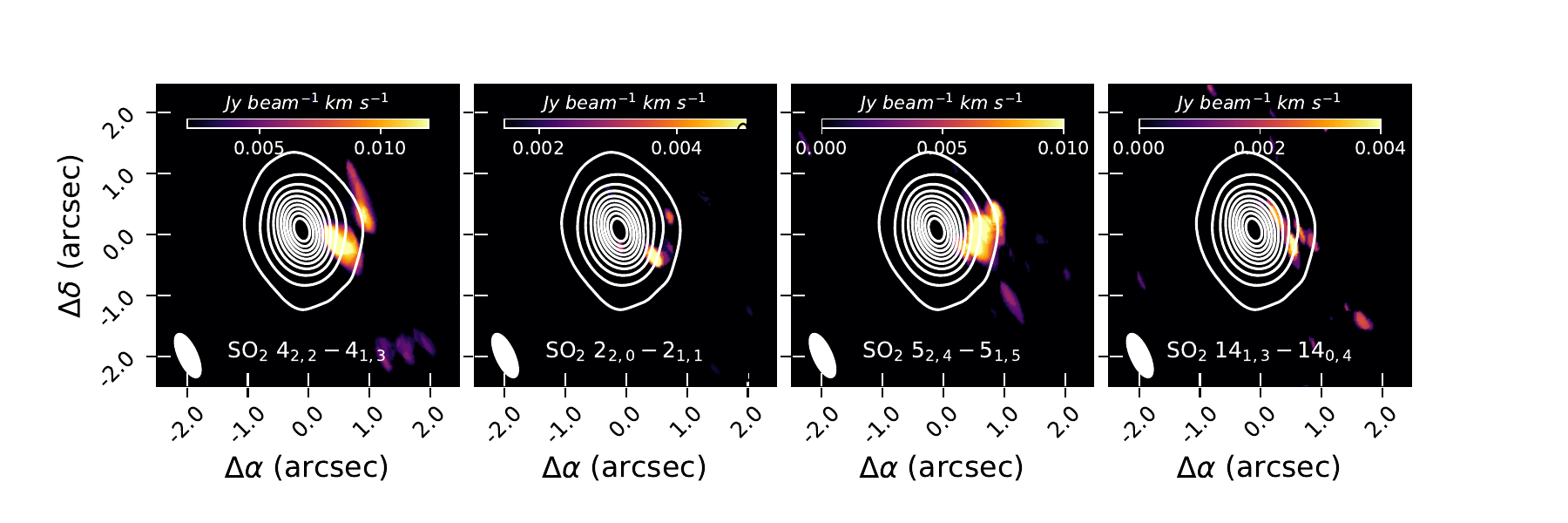}
  \caption{Zeroth-moment maps of the four SO$\rm _2$ transitions detected. The white overlayed contours show the distribution of continuum emission for comparison purposes. Continuum contours include ten levels that start at the 5$\sigma$ level and end at 0.9 of the emission peak. }
 \label{Fig:zeroth_moments_SO2}
\end{center}
\end{figure*}

We detected four SO$\rm _2$ transitions, 2$\rm _{2,0}$-2$\rm _{1,1}$, 4$\rm _{2,2}$-4$\rm _{1,3}$, 5$\rm _{2,4}$-5$\rm _{1,5}$, and 14$\rm _{1,13}$-14$\rm _{0,4}$, with upper energy levels in the range 12 to 101 K (see Table \ref{Tab:line_fluxes}). The non-detected SO$\rm _2$ transitions have energy levels ranging from 150 to 168 K. Figure \ref{Fig:zeroth_moments_SO2} shows the integrated intensity maps for these transitions. Emission is detected only on the western side. 

The 5$\rm _{2,4}$-5$\rm _{1,5}$ transition shows the strongest emission extending along the western side of the disk as well as along the extended filament observed in the CS map. The 4$\rm _{2,2}$-4$\rm _{1,3}$ and 14$\rm _{1,13}$-14$\rm _{0,4}$ transitions also show emission along this structure. The 2$\rm _{2,0}$-2$\rm _{1,1}$ transition, which has the most compact distribution, shows only a bright spot of emission in the southern part of the western side, and there is no emission along the position of the CS filament. \cite{Garufi2022} detected emission from 5$\rm _{2,4}$-4$\rm _{1,3}$, 14$\rm _{0,14}$-13$\rm _{1,13}$, and 3$\rm _{3,1}$-2$\rm _{2,0}$. Seven SO$\rm _2$ transitions have been detected in HL Tau so far.

\begin{figure}[t!]
\begin{center}
  \includegraphics[trim=5mm 0mm 0mm 0mm, clip, width=0.45\textwidth]{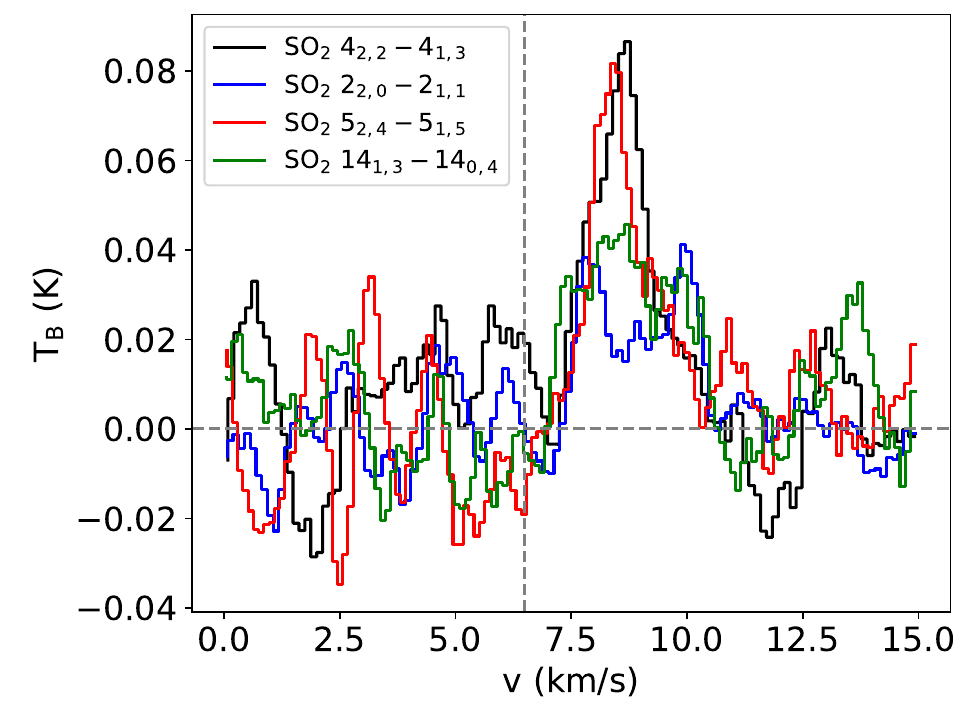}\\
  \caption{Continuum-subtracted average spectra of the detected SO$\rm _2$ transitions. The spectra shown have been integrated within a radius of 2$\arcsec$ from the center on the western side of the disk. The vertical dashed gray line depicts the position of maximal absorption in the CS map.}
 \label{Fig:SO2_integrated_spectra}
\end{center}
\end{figure}

In Fig. \ref{Fig:SO2_integrated_spectra}, we show the integrated spectra of the detected SO$\rm _2$ transitions. The spectra are in emission mostly from 7.5 to 10 km s$\rm ^{-1}$, compared to CS, H$\rm _2$S, and H$\rm _2CO$, where emission is observed from roughly 2.5 km s$\rm ^{-1}$ to 12.5 km s$\rm ^{-1}$, i.e., a range of 10 km s$\rm ^{-1}$ compared to 2.5 km s$\rm ^{-1}$ in the SO$\rm _2$ spectra. This is because SO$\rm _2$ emission is only detected on the western side of the disk. 

Table \ref{Tab:line_fluxes} includes line fluxes derived from integrated intensity maps. We integrated inside the beam area of the 4$\rm _{2,2}$-4$\rm _{1,3}$ transition centered at the position of the peak of the same transition line, following the beam-integrated method used by \cite{Garufi2022}. To ensure that our flux was compatible with the one used in \cite{Garufi2022}, we recomputed the fluxes for the transitions detected in that work and retrieved compatible values.

\begin{figure}[t!]
\begin{center}
  \includegraphics[trim=0mm 0mm 0mm 0mm, clip, width=0.45\textwidth]{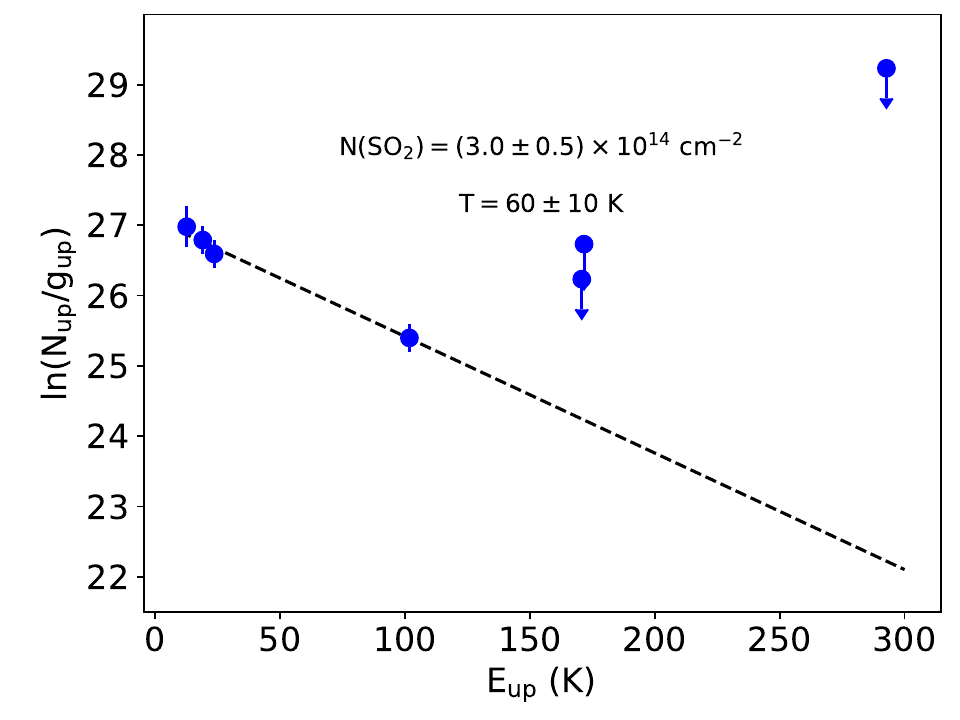}\\
  \includegraphics[trim=0mm 0mm 0mm 0mm, clip, width=0.45\textwidth]{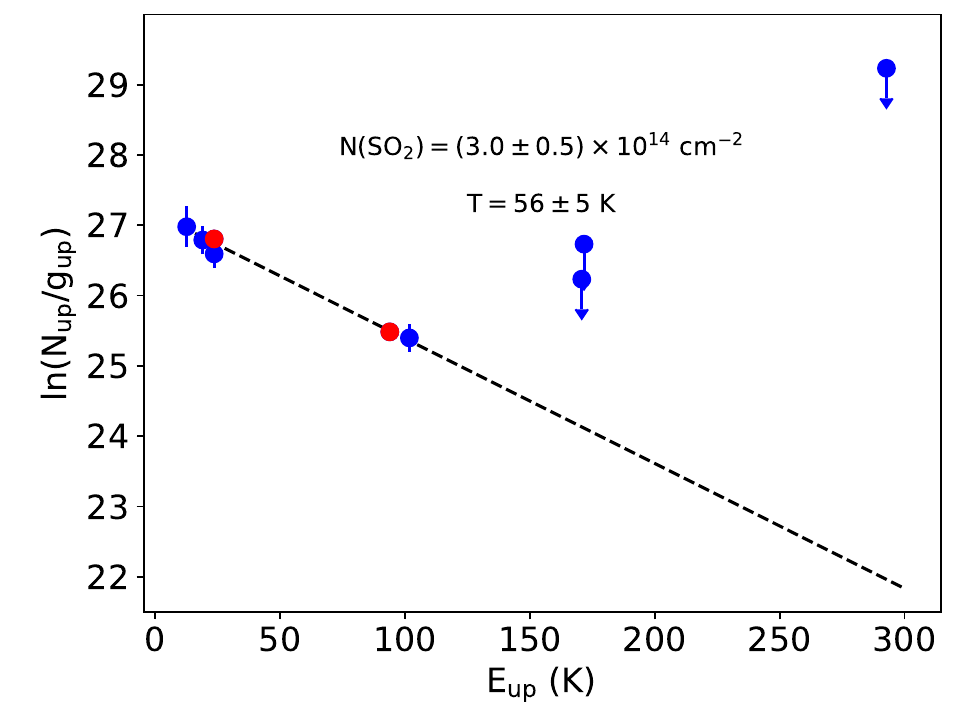}
  \caption{Rotational diagram of SO$\rm _2$ transitions detected with NOEMA. The top panel shows the rotational diagram for the transitions reported in this paper alone, while the bottom panel also includes the detections by \cite{Garufi2022} in red. }
 \label{Fig:SO2_rot_diag}
\end{center}
\end{figure}

We used the detected transitions to compute the rotational diagram of SO$\rm _2$. Assuming optically thin emission and the Rayleigh-Jeans limit, the rotation temperature and molecular column density can be derived from the equations

\begin{equation}
ln\left({N_{up}}\over{g_{up}}\right) = ln\left({N_{tot}}\over{Q_{rot}}\right) - \frac{E_{up}}{kT_{rot}}
\end{equation}
\begin{equation}
{\rm N}_{up} (cm^{-2})= 1.94 \times 10^3 \nu^2 (GHz) {\rm W} / {\rm A_{ul}} (s^{-1}) 
,\end{equation}
where $W$ is the velocity-integrated line area in units of Kelvins kilometer per second, $\rm Q_{rot}$ is the rotational partition function, $\rm E_{up}$ is the energy of the upper level, $\rm A_{ul}$ is the Einstein coefficient for spontaneous emission, and $\nu$ is the frequency of the transition. The partition function was computed for 198 rotational levels. The results are shown in Fig. \ref{Fig:SO2_rot_diag}. We derived a temperature of (60$\rm \pm$10) K and a column density of (3.0$\rm \pm$0.5)$\rm \times 10^{14}~cm^{-2}$. The temperature is compatible with T$\rm _{rot}=(58 \pm 19 )~K$ derived by \cite{Garufi2022}, but our column density is two times smaller than the value reported by them. We repeated the analysis, including the detections by \cite{Garufi2022}. We retrieved a flux of (120$\rm \pm$8) mJy km s$\rm ^{-1}$ for the 5$\rm _{2,4}$-4$\rm _{1,3}$ transition and of (164$\rm \pm$8) mJy km s$\rm ^{-1}$ for the 14$\rm _{0,14}$-14$\rm _{1,13}$ transition. This resulted in a temperature of (56$\rm \pm$5) K and a column density of (3.0$\rm \pm$0.3)$\rm \times 10^{14}~cm^{-2}$. We show the rotational diagram including the observations by \cite{Garufi2022} in the bottom panel of Fig. \ref{Fig:SO2_rot_diag}. Since we cannot rule out optically thick SO$\rm _2$ emission lines, the column densities reported here are lower limits. Figure \ref{Fig:SO2_rot_diag} also includes upper limits for the non-detected transitions with higher E$\rm _{up}$, but they do not constrain the temperature and column density. 
 

\begin{table*}[]
\caption{Column densities and molecular ratios in the disk and envelope.}
\label{Tab:col_dens}
\centering
\begin{tabular}{llllccc}
\hline \hline
Position & N(CS) & N(H$\rm _2$S) &  N(H$\rm _2$CO) &  N(CS)/N(H$\rm _2$S) & N(CS)/N(H$\rm _2$CO)  & N(H$\rm _2$S)/N(H$\rm _2$CO) \\
(--) & (cm$\rm ^{-2}$) & (cm$\rm ^{-2}$) & (cm$\rm ^{-2}$) \\
\hline
\hline
\multicolumn{7}{c}{T = 20 K} \\
CS peak & (3.6$\rm \pm$0.6)$\rm \times 10^{12}$ & (5.3$\rm \pm$1.2)$\rm \times 10^{12}$ &(4.1$\rm \pm$0.8)$\rm \times 10^{12}$ & 0.7$\rm \pm$0.3 & 0.9$\rm \pm$0.3 & 1.3$\rm \pm$0.5 \\
H$\rm _2$S peak & (2.9$\rm \pm$0.6)$\rm \times 10^{12}$ & (7.3$\rm \pm$1.2)$\rm \times 10^{12}$ & (3.6$\rm \pm$0.8)$\rm \times 10^{12}$ & 0.4$\rm \pm$0.1 & 0.8$\rm \pm$0.3 & 1.3$\rm \pm$0.4 \\
H$\rm _2$CO peak & (3.3$\rm \pm$0.6)$\rm \times 10^{12}$ & (2.9$\rm \pm$1.1)$\rm \times 10^{12}$ & (5.8$\rm \pm$0.8)$\rm \times 10^{12}$ & 1.1$\rm \pm$0.7 & 0.6$\rm \pm$0.2 & 0.5$\rm \pm$0.3 \\
Disk mean &(1.8$\rm \pm 0.8$)$\rm \times 10^{12}$ & (2.9$\rm \pm 1.3$)$\rm \times 10^{12}$ & (2.8$\rm \pm$1.1)$\rm \times 10^{12}$ & 0.7$\rm \pm$0.3 & 0.6$\rm\pm$0.2 & 1.0$\rm \pm$0.4 \\
\multicolumn{7}{c}{T = 40 K} \\
CS peak & (5.0$\rm \pm$0.9)$\rm \times 10^{12}$ & (7.5$\rm \pm$1.6)$\rm \times 10^{12}$ &(8.2$\rm \pm$1.6)$\rm \times 10^{12}$ & 0.7$\rm \pm$0.3 & 0.6$\rm \pm$0.2 & 0.9$\rm \pm$0.4 \\
H$\rm _2$S peak & (4.0$\rm \pm$0.9)$\rm \times 10^{12}$ & (1.0$\rm \pm$0.2)$\rm \times 10^{13}$ & (7.2$\rm \pm$1.6)$\rm \times 10^{12}$ & 0.4$\rm \pm$0.2 & 0.6$\rm \pm$0.2 & 1.4$\rm \pm$0.5 \\
H$\rm _2$CO peak & (4.7$\rm \pm$0.9)$\rm \times 10^{12}$ & (4.0$\rm \pm$1.6)$\rm \times 10^{12}$ & (1.1$\rm \pm$0.2)$\rm \times 10^{13}$ & 1.2$\rm \pm$0.7 & 0.4$\rm \pm$0.1 & 0.3$\rm \pm$0.2 \\
Disk mean &(2.5$\rm \pm 1.1$)$\rm \times 10^{12}$ & (4.0$\rm \pm 1.8$)$\rm \times 10^{12}$ & (5.5$\rm \pm$2.3)$\rm \times 10^{12}$ & 0.7$\rm \pm$0.3 & 0.4$\rm\pm$0.1 & 0.7$\rm \pm$0.3 \\
\multicolumn{7}{c}{T = 58 K} \\
CS peak & (6.5$\rm \pm$0.6)$\rm \times 10^{12}$ & (1.03$\rm \pm$0.06)$\rm \times 10^{13}$ &(1.25$\rm \pm$0.04)$\rm \times 10^{13}$ & 0.6$\rm \pm$0.2 & 0.5$\rm \pm$0.2 & 0.8$\rm \pm$0.3 \\
H$\rm _2$S peak & (5.2$\rm \pm$0.6)$\rm \times 10^{12}$ & (1.41$\rm \pm$0.06)$\rm \times 10^{13}$ & (1.11$\rm \pm$0.04)$\rm \times 10^{13}$ & 0.4$\rm \pm$0.1 & 0.5$\rm \pm$0.2 & 1.3$\rm \pm$0.4 \\
H$\rm _2$CO peak & (6.0$\rm \pm$0.6)$\rm \times 10^{12}$ & (5.5$\rm \pm$0.6)$\rm \times 10^{12}$ & (1.76$\rm \pm$0.04)$\rm \times 10^{13}$ & 1.1$\rm \pm$0.4 & 0.3$\rm \pm$0.2 & 0.3$\rm \pm$0.1 \\
Disk mean &(3.2$\rm \pm 1.4$)$\rm \times 10^{12}$ & (5.6$\rm \pm 2.6$)$\rm \times 10^{12}$ & (8.4$\rm \pm$4.0)$\rm \times 10^{12}$ & 0.7$\rm \pm$0.2 & 0.4$\rm\pm$0.1 & 0.6$\rm \pm$0.2 \\
\hline
\multicolumn{7}{c}{$\rm T_{ex}$ = 5 K} \\
Envelope & (4.7$\rm \pm0.4$)$\rm \times 10^{14}$ &  (3$\rm \pm$1)$\rm \times 10^{13}$ &  (2.0$\rm \pm$0.9)$\rm \times 10^{14}$ & 28$\rm \pm$12 & 4$\rm \pm$2 & 0.15$\rm \pm$0.12 \\
\multicolumn{7}{c}{$\rm T_{ex}~\to$ 0 K} \\
 & (2.7$\rm \pm0.6$)$\rm \times 10^{14}$ &  (1.3$\rm \pm$0.4)$\rm \times 10^{13}$ &  (7$\rm \pm$3)$\rm \times 10^{13}$ & 16$\rm \pm$8 & 3$\rm \pm$2 & 0.19$\rm \pm$0.14 \\
\hline
\end{tabular}
\tablefoot{The uncertainty in the column density at the different positions is computed as the quadratic sum of the propagated rms and the propagated calibration uncertainty (10\%). The uncertainty in the mean value is the standard deviation. }
\end{table*}

\subsection{Molecular column densities}
\begin{figure}[t!]
\begin{center}
  \includegraphics[width=0.45\textwidth]{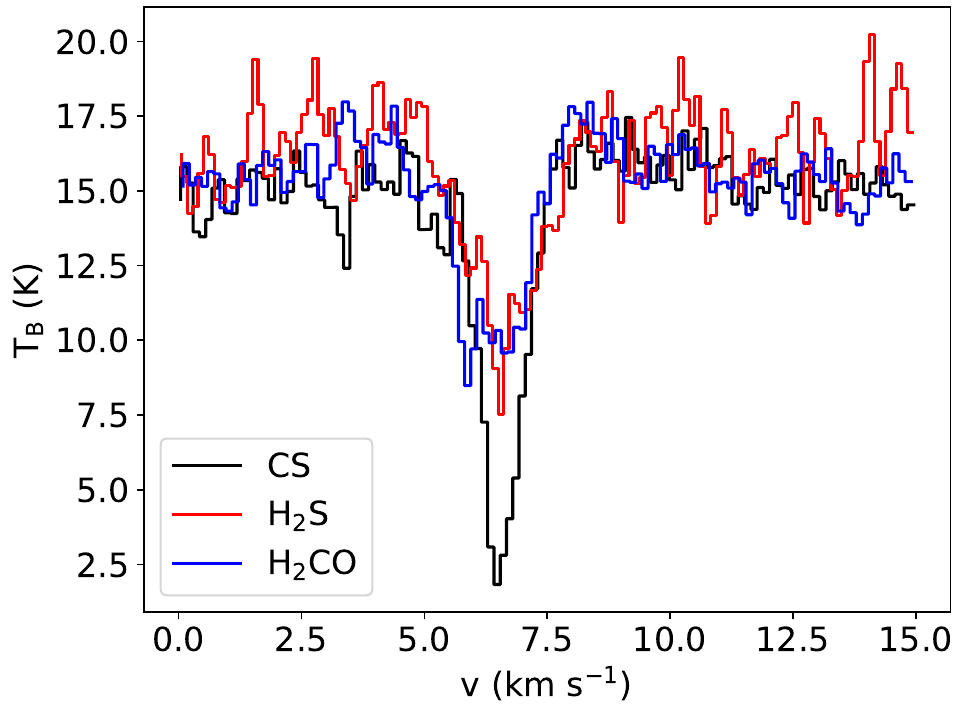}\\
  \caption{Observed spectra at the position of maximum absorption. No continuum was subtracted.}
 \label{Fig:peak_abs_spec}
\end{center}
\end{figure}

\begin{figure}[t!]
\begin{center}
  \includegraphics[trim=0mm 0mm 0mm 0mm, clip, width=0.45\textwidth]{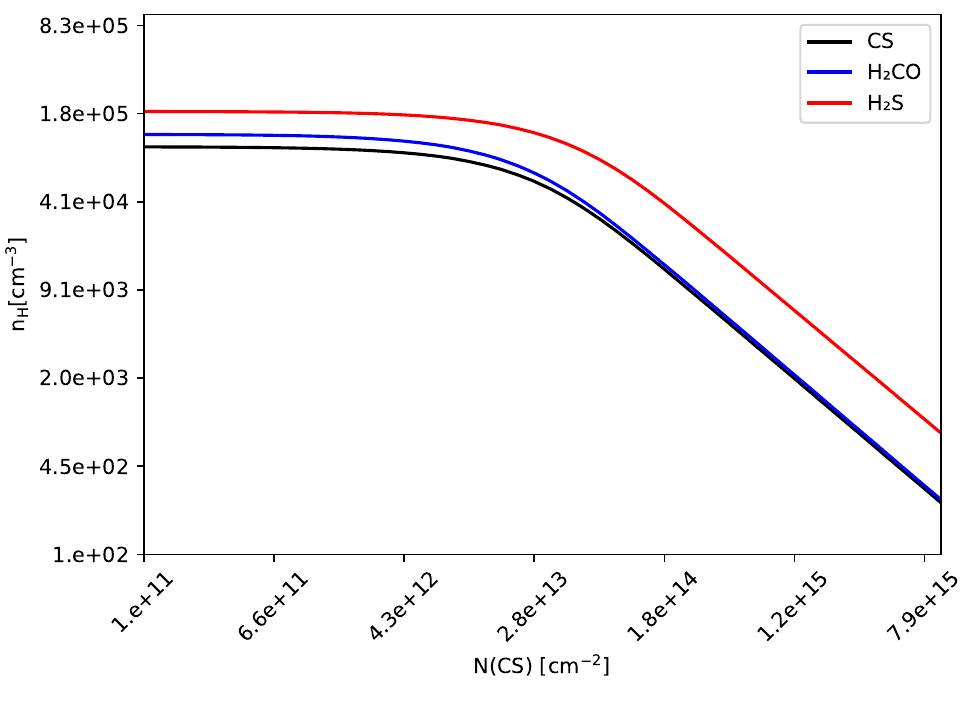}\\
  \caption{Excitation temperature, T$\rm _{ex}$, of $5~K$ as a function of n$\rm _H$ and N(X) for CS (black), H$\rm _2$S (red), and H$\rm _2$CO (blue) from RADEX models. The black, blue, and red lines mark the position of the $\rm T_{ex}=5~K$ contour for each species.}
 \label{Fig:Tex_RADEX_grid}
\end{center}
\end{figure}

We used the absorption features toward the center to derive the column density of the species surveyed along the line of sight. We selected a single position toward the absorption peak of CS. To estimate the column densities, we assumed that the source fills the beam. Given the assumption, we could use

\begin{equation}
\tau = -\ln \left(\frac{T_B-J_\nu(T_{ex})}{T_{BG}-J_\nu(T_{ex})}\right),
\end{equation}

\noindent
where $T_B$ is the measured brightness temperature (line plus continuum), $T_{BG}$ is the background temperature derived from continuum-only maps, and 

\begin{equation}\label{eq:J_nu}
J_\nu(T_{ex}) = \frac{h\nu/k}{e^{h\nu/KT_{ex}}-1}.
\end{equation}
Assuming that the continuum source fills the beam, Eq. \ref{eq:J_nu} constrains $\rm T_{ex} \le 5~K$, since we would not see absorption features otherwise due to the low temperatures measured in the line emission maps. At the position of maximum absorption, the line plus continuum map for CS reaches a temperature of 2.5 K (see Fig. \ref{Fig:peak_abs_spec}), and hence we needed $\rm J_{146GHz}(T_{ex}) < 2.5 K$. For an excitation temperature of 5.4 K,  $\rm J_{146GHz}(5.4) = 2.6 K$, but it goes down to 2.3 at 5 K. Thus, we assumed an excitation temperature of $\rm T_{ex} = 5~K$.  To further support our assumption regarding T$\rm _{ex}$, we computed a grid of RADEX models \citep{vanDerTak2007} for a set of densities and CS, H$\rm_2$S, and H$\rm _2$CO column densities resembling those in a cold protostellar envelope while assuming a gas temperature of 15 K in the envelope \citep{Yen2017}.  We show the resulting models in Fig. \ref{Fig:Tex_RADEX_grid}. For models with N(X)$\rm <$10$\rm ^{13}~cm^{-2}$ and $\rm n_{H_2} <10^5~cm^{-3}$, T$\rm _{ex}$ is always below 5 K. \cite{Wu2018} modeled the volume density using the equation 
\begin{equation}
n_{H_2} (r, |z|) = \frac{(2-\gamma)M_d}{2 \pi r_c^2 \sqrt{2 \pi} H(r)m_H} \left( \frac{r}{r_c} \right)^{-\gamma}e^{-\frac{ |z|^2}{2H^2(r)}},
\end{equation}
where $\rm M_d$ is the disk mass, $\rm r_c=100 ~au$ is the reference radius, H(r) is the radius-dependent scale height, $\rm m_H$ is the hydrogen mass, and $\gamma$ is the power-law index of the volume density profile, $\gamma$=0.5 in the envelope. This equation results in  $\rm n_{H_2}$ $\rm < 10^{3}~cm^{-3}$ at z$\rm \sim$1000 au. Given the inclination of the disk and the envelope, the line of sight toward the absorption feature should only pass through the upper layers of the envelope where  $\rm n_{H_2}$ is low. To test the impact of the assumed excitation temperature, we also computed the column densities when $\rm T_{ex} \to 0$. Since $\lim_{T_{ex} \to 0}J_\nu=0$, we used

\begin{equation}
\tau = -\ln \left(\frac{T_B}{T_{BG}}\right).
\end{equation}

The column density can be derived from the $\tau$ profile using

\begin{equation}
N_{tot} = \frac{3h}{8 \pi^3 |\mu|_{lu}^2} \frac{Q_{rot}}{g_u} e^{{\frac{E_u}{kT_{ex}}} } \left[  e^{\frac{h\nu}{kT_{ex}}}-1 \right]^{-1} \int \tau_\nu dv.
\end{equation}
We applied the equations mentioned above to the absorption features in our NOEMA maps and derived the column density of CS, H$\rm _2$CO, and H$\rm _2$S in the absorbing envelope. Molecular parameters were retrieved from CDMS. We computed the average column densities by averaging the derived column densities of pixels inside the area showing absorption features. To compute the total column densities for H$\rm_2$S and H$\rm _2$CO, we have to correct for the ortho-to-para ratio. Assuming that H$\rm _2$S behaves similar to H$\rm _2$O, we used an ortho-to-para ratio of three \citep{Hama2016}. The same value was used for H$\rm _2$CO since laboratory experiments show a narrow range of ortho-to-para ratios, 2.9 to 3.0, for temperatures characteristic of the warm molecular layer \citep{Yocum2023}. Thus, we multiplied the column densities computed for the o-species by a factor of 1.33 to retrieve the total column densities included in Table \ref{Tab:col_dens}. The column densities for $\rm T_{ex}=5~K$ act as upper limits, while the values computed when $\rm T_{ex} \to 0K$ serve as lower limits.

\begin{figure}[t!]
\begin{center}
  \includegraphics[trim=0mm 0mm 0mm 0mm, clip, width=0.45\textwidth]{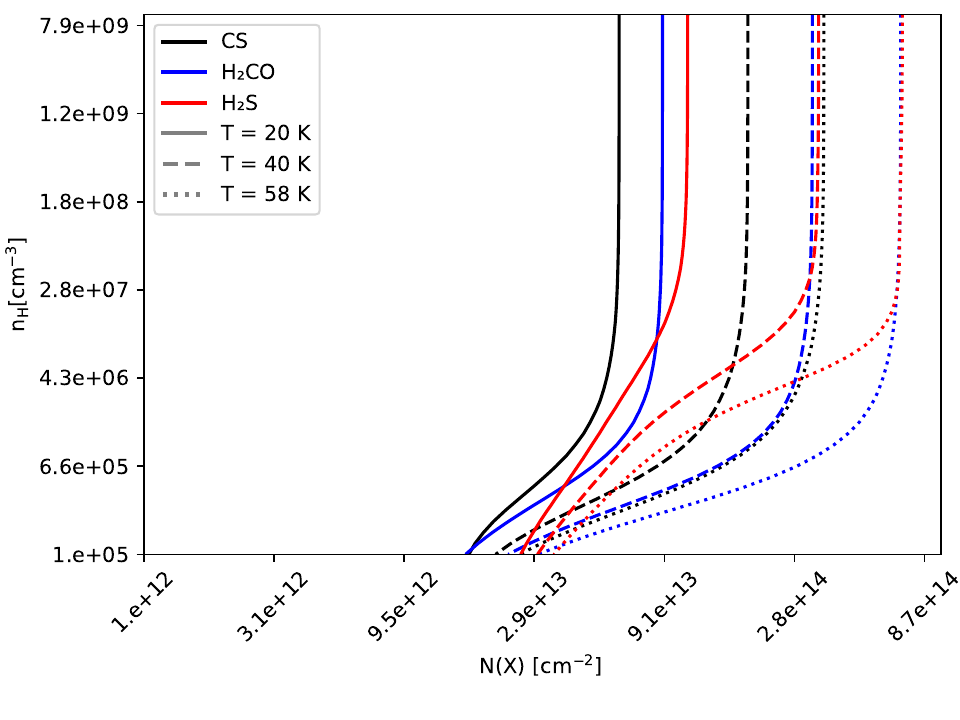}\\
  \caption{Excitation temperature, T$\rm _{ex}$, of $5~K$ as a function of n$\rm _H$ and N(X) for CS (black), H$\rm _2$S (red), and H$\rm _2$CO (blue) from RADEX models. The white line marks the position of the $\rm T_{ex}=5~K$ contour.}
 \label{Fig:RADEX_grid_Tau}
\end{center}
\end{figure}

Since we are interested in comparing the column density ratios of the detected species in the disk and envelope, we also computed the column densities in the disk. We only have one transition per molecule, so we assumed local thermodynamical equilibrium (LTE) and optically thin emission. \cite{Wu2018} assumed LTE to perform radiative transfer calculations based on the $\rm ^{13}CO(2-1)$/$\rm ^{13}CO(1-0)$ ratio. Molecular emission in HL Tau most likely comes from the warm molecular layer. We derived critical densities for the studied transitions using the Einstein coefficients for spontaneous emission from Table \ref{Tab:line_fluxes} and collisional rates from \cite{DenisAlpizar2013} for CS, \cite{Dagdigian2020} for H$\rm _2$S, and \cite{Wiesenfeld2013} for H$\rm _2$CO. The critical densities for the observed transitions are in the range from 4$\rm \times 10^{5}~cm^{-3}$ to $\rm 10^{6}~cm^{-3}$ for temperatures in the range from 20 to 60 K. Thus, at the high densities that are found in the warm molecular layer \citep[$\rm 10^6$-$\rm 10^8~cm^{-3}$][]{vanZadelhoff2001}, the studied species will be in or close to LTE.  In \cite{Garufi2022}, a temperature of 58$\rm \pm$19 K was derived for SO and SO$\rm _2$ using rotational diagrams. Using our NOEMA observations and ALMA, we derived a temperature of (56$\rm \pm$5) K, which is the temperature that we assumed to compute column densities. Using RADMC-3D, \cite{Jin2016} produced a hydrodynamic simulation of the gas and dust in HL Tau and derived surface temperatures ranging from 280 K at 5 au to 75 K at 100 au and mid-plane temperatures ranging from 110 K at 5 au to 20 K at 100 au. \cite{Pinte2016} used MCFOST to model self-consistently dust emission in HL Tau and obtained the same mid-plane temperature at 100 au, 20 K. The temperature profile assumed by \cite{Okuzumi2016} results in T = 22 K at 100 au, and the same temperature was derived by \cite{Carrasco2016} using ALMA and Very Large Array observations. \cite{Liu2017} performed detailed radiative transfer modeling of the dust emission in the HL Tau protoplanetary disk and derived a slightly larger mid-plane temperature of 25 K at 100 au. Thus, to provide an estimate of the uncertainty associated with our temperature selection, we also computed column densities at 20 K (the mid-plane dust temperature, which can be considered as the lower limit) and 40 K. We calculated column densities in three different positions: that of the CS peak, that of the H$\rm _2$CO peak, and that of the H$\rm _2$S peak. We note that the western side is contaminated with emissions from the streamer and/or the spiral arm. We provide a summary of molecular column densities in Table  \ref{Tab:col_dens}, where we also include molecular ratios to help with the comparison. We also note that although molecular column densities vary depending on the assumed gas temperature, the ratios remain mostly constant, and the differences observed between the envelope and the protoplanetary disk are a robust result.

The optical thickness of the HL Tau dust in the outer disk at 2mm is $\tau$ $\sim1$ \citep{Carrasco2019}, which can result in continuum oversubtraction and suppressed line intensities. However, continuum oversubtraction also requires optically thick line emission. We show in Fig. \ref{Fig:RADEX_grid_Tau} the curve for $\tau$ = 1 for a series of RADEX models as a function of the number density, $\rm n_{H_2}$, and the column density, N(X), for three different kinetic temperatures, 20 K, 40 K, and 58 K. We observed that at 58 K and for the range of $\rm n_{H_2}$ that characterizes the warm molecular layer, the line emission of the surveyed species will be optically thick only for large column densities, which are not expected in protoplanetary disks. However, if the emission originates in a region with lower temperatures, such as the mid-plane, over-subtraction could be an issue. In any case, the reported column densities act as lower limits.

\section{Discussion}\label{Sect:discussion}

We have presented new NOEMA observations of CS, H$\rm _2$S, H$\rm _2$CO, and SO$\rm _2$. We find that CS, H$\rm _2$S, and H$\rm _2$CO show similar spatial distributions, with radial peaks at almost the same distance from the central star ($\rm \sim$105 au). Observations of H$\rm _2$S and H$\rm _2$CO in AB Aur also resulted in similar radial profiles \citep{Riviere2022}. Regarding azimuthal variations, the three species show different distributions: CS and H$\rm _2$S peaks are on the western side of the disk, while H$\rm _2$CO peaks at the eastern side (see Fig. \ref{Fig:moments_KeplerMask}). SO$\rm _2$ emission is only detected on the western side and is likely due to desorption from dust grains after the southern streamer traced by CS impacts the protoplanetary disk. The streamer's location is well traced by the HCO$\rm ^+$ 3-2 emission \citep{Yen2019} as well as by CS 3-2 (Fig. \ref{Fig:moments}). Furthermore, the SO$\rm _2$ maps show faint emission at the position of the streamer. We did not detect non-Keplerian motions at the position of the streamer in our Keplerian masked integrated intensity maps. However, \cite{Garufi2022}, using simple models for the velocity field, detected prominent non-Keplerian residuals, which they attributed to the impact of the streamer. 

\subsection{Column densities and molecular ratios: Envelope versus disk}\label{Subsect:cold_dens_and_ratios}
Table \ref{Tab:col_dens} lists the column densities derived in this study for H$\rm _2$CO, H$\rm _2$S, and CS in the disk and in the envelope. The CS, H$\rm _2$S, and $\rm H_2CO$ estimated column densities in the disk increase on average by 40\%, 39\%, and 97\%, respectively, when the adopted excitation temperature goes from 20 to 40 K, and by 81\%, 92\%, and 205\% when the temperature goes from 20 to 58 K. Thus, within the temperature range explored, the species that is most affected by a change in temperature is H$\rm _2$CO. This is an expected result since H$\rm _2$CO has long been known to be a good tracer of temperature \citep{Mangum1993,Muhle2007,Tang2017}. It is worth mentioning, however, that the differences with temperature can be more dependent on the observed transitions than on the species themselves. The mean molecular ratios computed in the disk are in stark contrast with the ratios derived for the envelope (see Table \ref{Tab:col_dens}). The most notable difference is in the N(CS)/N(H$\rm _2$S) ratio, which is between 23 and 70 times larger in the envelope. This suggests chemical evolution in the transition from the envelope to the disk. Furthermore, the difference in the computed ratios highlights the potential of molecular ratios, especially that of N(CS)/N(H$_2$S), as tools for differentiating envelopes and disks.

The differences between the envelope and the disk originate in the different excitation regimes between the quiescent and cold envelope and the warmer disk as well as in the different UV irradiation regimes. The disk is subject to strong UV irradiation from the central protostar, while the envelope, which is farther out, is less affected. The physical conditions are also different in the two components, as shown by \cite{Wu2018}. Specifically, \cite{Wu2018} performed a detailed modeling of a set of observations of gaseous species and the continuum, and they derived different exponents for the radial dependence, with $\gamma = 1$ in the disk and  $\gamma = 0.5$ in the envelope. The temperature in the disk is higher, thus resulting in different excitation conditions. The disk and envelope followed different evolutionary paths. Furthermore, the disk is also subject to shocking activity caused by the impact of large-scale streamers on the western side of the disk. The impact of the streamer can result in the sputtering of molecular species such as H$\rm _2CO$ and H$\rm_2$S from the surface of grains, while CS, which is formed in the gas phase, is not subject to local enrichment in the disk. The impact of the streamer generates a region of enhanced turbulence \citep{Kuznetsova2022} that can result in variations in the abundance of sensitive species \citep[][ see Sect. \ref{Subsect:mol_ratio_disk_variations}]{Semenov2011, Heinzeller2011}. \cite{Garufi2022} observed enhanced SO and SO$\rm _2$ emission in the region where the streamers impact the HL Tau and DG Tau disks, indicating a localized release of species linked to the impact of the streamer. The SO$\rm _2$ detections presented in this paper, which are only observed in the western side (where the streamer seems to impact the protoplanetary disk), further support this scenario. \cite{Tanious2024} also observed enhanced SO emission at the position where the streamer likely impacts the L 1489 IRS disk. \cite{Taniguchi2024} measured slightly higher temperatures and densities at the position where the streamer impacts the disk in their study of Per-emb-2. They concluded that these positions were also chemically younger. \cite{Bianchi2023} studied molecular emission in SVS 13-A and discarded thermal sublimation as the origin of the observed emission, concluding that the morphology was consistent with an accretion shock. The mechanism behind the origin of SO$\rm _2$ might also explain the enhancement of H$\rm _2$S and H$\rm _2$CO. Furthermore, if the impact of the streamer increases the temperature of the gas and dust \citep{Taniguchi2024}, it could also enrich the environment via thermal desorption from dust grains. H$\rm _2$CO desorbs from the ice at dust temperatures greater than 65 K \citep{vanGelder2021}, which is close to the value that we measured for the rotational temperature of SO$\rm _2$. Furthermore, the binding energies for H$\rm _2$S and H$\rm _2$CO are lower than that of SO$\rm _2$ \citep{Noble2012, Ferrero2020, Perrero2022}, so we expect H$\rm _2$S and H$\rm _2$CO to be effectively desorbed in an environment where SO$\rm _2$ has been released from the surface of grains.

 The enhancement in the disk compared to the envelope of the amount of H$\rm _2$S relative to H$\rm _2$CO likely originates in the different photodesorption yields of the two species. According to \cite{Fuente2017}, the photodesorption yield for H$\rm _2$S is 1.2$\rm \times 10^{-3}$ molecules per incident photon, compared to 6$\rm \times 10^{-5}$ for H$\rm _2$CO \citep{Martin2016}, so the H$\rm _2$S photodesorption yield is approximately 20 times larger than the H$\rm _2$CO one.  Thus, H$\rm _2$S is more effectively photo-desorbed from the surface of dust grains in the disk than is H$\rm _2$CO, resulting in an enhancement of H$\rm _2$S relative to H$\rm _2$CO in the gas phase. 

\subsection{Molecular ratios: Disk variations}\label{Subsect:mol_ratio_disk_variations}
Molecular ratios are essential tools to study the physical conditions and chemical composition of interstellar environments. It is thus interesting to extend the comparison to molecular ratios at different positions within the disk to see to what extent the chemistry depends on its precise position. We tested three positions: the CS peak, the H$\rm _2$S peak, and the H$\rm _2$CO peak. We also tested three temperature regimes: 20, 40K, and 58 K \citep{Garufi2022}. While we noted differences in the mean values of the molecular ratios, they are compatible within the uncertainties. However, we observed statistically meaningful differences in the ratios computed at different positions in the disk. The most prominent difference is that of the N(H$\rm _2$S)/N(H$\rm _2$CO) ratio, which at 40 K is almost five times larger at the H$\rm _2$S peak compared to the H$\rm _2$CO peak. A similar range of variation was observed at 58 K. These differences in molecular ratios along the disk's azimuth indicate variations in the chemical evolution within the disk. Furthermore, we note that the SO$\rm _2$ emission comes from the western side of the disk alone, further indicating chemical differences with azimuth. The differences from the west side to the east side seem to be linked with the presence of a large-scale streamer that impacts the western side of the disk. 

The variations of this ratio observed along the disk might indicate regions with different levels of turbulence. \cite{Heinzeller2011} showed that turbulent mixing has an impact on the disk chemistry, altering the abundance of species in the warm molecular layer of protoplanetary disks. Sulfur-bearing species were among the most enhanced by turbulence. \cite{Semenov2011} classified CS as a species sensitive to turbulence, meaning that its abundance changed by up to two orders of magnitude when turbulent transport was included in the model, while H$\rm _2$S and H$\rm _2$CO were included in the steadfast group, meaning that their abundances varied by a factor of three to five. \cite{Semenov2011} noted that abundances were already affected by a low Schmidt number of 1, while a Schmidt number of 100 resulted in larger variations for the sensitive group of molecules.

\subsection{SO$\rm _2$ detection}
While models indicate that SO$\rm _2$ is a major sulfur carrier in the ice surface of dust grains \citep{Riviere2022}, gas-phase SO$\rm _2$ has only been detected in a few sources \citep{Semenov2018, Booth2021, Garufi2022, Booth2024}. Its ice counterpart remains undetected. \cite{Garufi2022} reported the detection of three SO$\rm _2$ transitions toward HL Tau, 3$\rm _{3,1}$-2$\rm _{2,0}$, 5$\rm _{2,4}$-4$\rm _{1,3}$, and 14$\rm _{0,14}$-13$\rm _{1,13}$. Using NOEMA observations, we report the detection of another four SO$\rm _2$ transitions in HL Tau. The emission originates in the western side of the disk and is cospatial with the transitions mapped by \cite{Garufi2022}. This is the region where a large-scale streamer impacts the disk \citep{Yen2019, Garufi2022}. We did not detect emission at systemic velocities compatible with an origin in the central source. We speculate that the SO$\rm _2$ emission comes from gas desorbed from the icy surface of dust grains by shocks, similar to the other SO$\rm _2$ lines discussed by \citet[see our Sect. \ref{Subsect:cold_dens_and_ratios}]{Garufi2022}. The impact of the streamer in the disk results in a local enhancement of molecular species (including SO$\rm _2$) that are sputtered from the surface of dust grains in the protoplanetary disk. As pointed out by \cite{Garufi2022}, thermal desorption would result in an azimuthally symmetrical emission, rather than a localized emission. Using rotational diagrams, we derived a temperature of 56 K and an SO$\rm _2$ column density of 3$\rm \times 10^{14}~cm^{-2}$. Our temperature is in good agreement with the value derived by \citet[58$\rm \pm$19]{Garufi2022} K, but the column density is less than half the value derived by them. Our estimate is, however, based on a larger number of measurements and thus is more robust. Since SO$\rm _2$ is only present in the western side, where the large-scale streamer impacts the disk, the computed SO$\rm _2$ column density traces desorbed material at $\rm \sim$56 K. The detection of SO$\rm _2$ emission toward the center by \cite{Garufi2022} points to the existence of two SO$\rm _2$ reservoirs: a cold asymmetric reservoir to the southwest of the disk, which likely results from the impact of the streamer on the disk, and a warm one in the innermost region observed in the 14$\rm _{0,14}$-13$\rm _{1,13}$ transition. This reservoir spatially overlaps with the H$\rm _2$O emitting region detected by \cite{Facchini2024}. We did not detect SO$\rm _2$ emission from the warm reservoir.

\section{Summary and conclusions}\label{Sect:summary}

We have presented new NOEMA interferometric observations toward HL Tau, which resulted in the detection of H$\rm _2$S, CS, H$\rm _2CO$, and SO$\rm _2$. For SO$\rm _2$, four transitions were imaged, allowing us to derive the gas column density and temperature by means of rotational diagrams. In the following, we summarize our main results:

\begin{itemize}
\item Three species, H$\rm _2$S, CS, and H$\rm _2CO$, are detected at all azimuths, while SO$\rm _2$ is only detected at the southwestern side of the disk. H$\rm _2$S, CS, and H$\rm _2CO$ also show absorption by the envelope toward the position of the central protostar. SO$\rm _2$ is not observed in absorption. H$\rm _2$S, CS, and H$\rm _2CO$ show radial profiles that peak at similar distances from the center ($\sim$105 au), which is in stark contrast with other protoplanetary disks such as AB Aur, but show azimuthal variations. 

\item We computed molecular column densities and ratios for the detected species in the disk and the envelope and observed strong variations in the ratios. The most prominent difference is in the N(CS)/N(H$\rm _2$S) ratio, which is 40 to 50 times larger in the envelope. This indicates that molecular ratios can be used to distinguish the emission coming from the protoplanetary disk from that coming from the protostellar envelope. Furthermore, the observed ratios suggest chemical evolution in the transition from the cold envelope to the protoplanetary disk. 

\item For SO$\rm _2$ we derived column densities and gas temperatures using rotational diagrams. We derive a temperature in good agreement with previous values in the literature, but the column density is roughly two times smaller. However, our estimates are better constrained due to the larger number of data points used. 

\item Differences in the molecular ratio can be seen not only when comparing the protoplanetary disk and the envelope but also when observing the distribution along the disk itself. Observations of a larger sample of species at a higher angular resolution are needed to study the chemical budget of HL Tau in detail, and its spatial distribution, and to understand the chemical differences between the envelope and the disk.
\end{itemize}

\section*{Data availability}
Data cubes are available at the CDS via anonymous ftp to cdsarc.u-strasbg.fr (130.79.128.5) or via http://cdsweb.u-strasbg.fr/cgi-bin/qcat?J/A+A/.

\begin{acknowledgements}
We want to thank the anonymous referee for a fruitful scientific discussion that has resulted in a much-improved manuscript. A.F., G.E., and P.R.M. are members of project PID2022-137980NB-I00, funded by MCIN/AEI/10.13039/501100011033/FEDER UE. I00.This project is co-funded by the European Union (ERC, SUL4LIFE, grant agreement No101096293). D.S. has received funding from the European Research Council (ERC) under the European Union's Horizon 2020 research and innovation programme (PROTOPLANETS, grant agreement No. 101002188; PI: M.~Benisty). This work is based on observations carried out under project number W22BA with the IRAM NOEMA Interferometer. IRAM is supported by INSU/CNRS (France), MPG (Germany) and IGN (Spain). This paper makes use of the following ALMA data: ADS/JAO.ALMA\#2018.1.01037.S. ALMA is a partnership of ESO (representing its member states), NSF (USA) and NINS (Japan), together with NRC (Canada), NSTC and ASIAA (Taiwan), and KASI (Republic of Korea), in cooperation with the Republic of Chile. The Joint ALMA Observatory is operated by ESO, AUI/NRAO and NAOJ. S.F. acknowledges financial contributions by the European Union (ERC, UNVEIL, 101076613), and from PRIN-MUR 2022YP5ACE. Views and opinions are however those of the author(s) only and do not necessarily reflect those of the European Union or the European Research Council. Neither the European Union nor the granting authority can be held responsible for them.
\end{acknowledgements}

 \bibliographystyle{aa} 
\bibliography{biblio}

\end{document}